\documentclass[twocolumn]{autart}    

\usepackage{graphicx,amsmath,amssymb,overpic,here,multirow}          

\usepackage{booktabs,bm}

\newtheorem{definition}{\bf Definition}[section]

\newtheorem{hyp}{\bf Assumption}

\usepackage[dvipsnames]{xcolor}
\usepackage{tikz}
\usepackage{pgfplots}
\usetikzlibrary{positioning}

\usepackage{algorithm}
\usepackage{algpseudocode}
\begin{document}

\begin{frontmatter}

\title{Explicit approximation of stochastic optimal feedback control for combined therapy of cancer\thanksref{footnoteinfo}} 

\thanks[footnoteinfo]{This paper was not presented at any IFAC meeting. Corresponding
author M.~Alamir. Tel. +33476826326. Fax +33476826388.}
\author[GIPSA]{Mazen Alamir}\ead{mazen.alamir@grenoble-inp.fr}    
\address[GIPSA]{Univ. Grenoble Alpes, CNRS, Grenoble INP, GIPSA-lab, 38000 Grenoble, France}
\begin{abstract}
In this paper, a  tractable methodology is proposed to approximate stochastic optimal feedback treatment in the context of mixed immuno-chemo therapy of cancer. The method uses a fixed-point value iteration that approximately solves a stochastic dynamic programming-like equation. It is in particular shown that the introduction of a variance-related penalty in the latter induces better results that cope with the consequences of softening the health safety constraints in the cost function. The convergence of the value function iteration is revisited in the presence of the variance related term. The implementation involves some Machine Learning tools in order to represent the optimal function and to perform complexity reduction by clustering. Quantitative illustration is given using a commonly used model of combined therapy involving twelve highly uncertain parameters.  
\end{abstract}
\end{frontmatter}

\section{Introduction}
Rationalizing drug delivery is an active research field that commonly involves population models with high number of uncertain parameters. Unfortunately, these parameters which are by nature highly variable between individuals, are inaccessible to identification because of the lack of excitation \cite{Kiran2010}. 

The great majority of applied mathematical-like works solve deterministic optimal control problems which are formulated using the  nominal values of the parameters \cite{Swan:88,DePillis06,Ledzewicz:08}. While this might be important to draw qualitative conclusions regarding the patterns of optimal strategies (intensive treatment, presence of singular arcs, etc); the resulting profiles do not accommodate for the high dispersion of the parameters.

Using repetitive solutions of such open-loop nominal scheduling strategies in a feedback mode through the receding-horizon principle (apply the first part of the optimal strategy and recompute a new optimal injection profile at the next decision instant and so on leading to the so called Model Predictive Control design (MPC).) obviously reduces the drawback of parameters mismatch \cite{Chareyron:09,CHEN2012973}. However, the performances can only be observed a posteriori as nothing is explicitly done to address the presence of uncertainties. 

Another option is to use a parameterized state feedback law and to optimally tune its design parameters using explicitly the statistical description of the model's parameters dispersion by means of the randomized optimization framework \cite{ALAMIR201559}. Unfortunately, as far as available works are concerned, this is to be done for each initial state in the proposed frameworks. 

Stochastic model predictive control (SMPC) offers an attractive alternative through the use of a set of scenarios in the on-line computation in order to minimize the approximate version of the expectation of the cost function \cite{Mesbah:2014,Mesbah:2018,cancer2018Arxiv}. This lead to a solution for each initial state but does not give a global overview on the performance on a whole region of the state space. On the other hand, this solution can be scalable in the dimension of the state which enables the use of almost arbitrarily complex models of the underlying dynamics. 

Stochastic Dynamic Programming (SDP) \cite{Jaakoola1994,Bertsekas2017,Kirkby2017} is a framework that, at least conceptually, outperforms the previous approaches in that it gives an 1) explicit state feedback law that 2) explicitly incorporates the statistical description of the parameters dispersion while being 3) defined on a whole region of interest within the state space. 

Unfortunately, the price to pay to get these advantages is to represent the Bellman function over the extended space of state and control making this approach {\bf non scalable} in the dimension of the state and control. Moreover, when minimizing the residual of the SDP equation, each call involves the computation of statistically defined quantities and this, for the computation to be relevant, should theoretically involve a high number of randomly sampled instances of the uncertain model's parameters. Note however that this last drawback is obviously shared by the SMPC approach. 

The literature on solving the stochastic dynamic programming equation is huge and a complete survey of it is out of the scope of this paper. For a complete and recent survey, one can consult \cite{Bertsekas2017}. This paper does not claim a particularly novel algorithm to solve the SDP equation. Rather it focuses on the issue of introducing a variance-related term in the definition of the cost function, investigates its impact on handling the health-related constraint in the combined therapy of cancer and revisits the convergence proof of the underlying value function fixed-point iteration when such additional variance-related term is added. 

More precisely, a simple framework is proposed for a moderate sized models\footnote{the combined therapy model used involves four states and two controls which are commonly encountered sizes in all the related works.} in which  Machine Learning (ML) tools are used. More precisely, the model's structure uses a Support Vector Machine (SVM) regression model to represent the value function. On the other hand, a clustering approach is used to approximate the statistical quantities using a lower number of samples. This leads to a tractable approximate solution that can be obtained in less than 20 minutes when using a grid of 9604 points in the state/control space when no parallelization is used. Drastic reduction of the computation time can be obtained since the proposed scheme is amenable to massive parallelization. 

This paper is organized as follows: The problem of combined therapy is stated in Section \ref{SecPs}. Some recalls regarding SDP are proposed in Section \ref{SecRecalls} and some results are proposed for the convergence of a fixed-point iteration that is used to solve the associated equations. Section \ref{SecPM} details the proposed approximate solution's framework while Section \ref{SecRes} shows the results on the combined therapy of cancer and discusses some issues. Finally, the paper ends with Section \ref{SecConc} that summarizes the paper's contributions and gives some hints for further investigation.   

\section{Problem statement} \label{SecPs}
\subsection{The dynamic model}
Let us consider the population model used in \cite{Kassara2011135} to describe the dynamics involved in the combined immuno-chemo therapy:
\begin{eqnarray}
\dot x_1&=&ax_1(1-bx_1)-c_1x_4x_1-k_3x_3x_1 \label{model1} \\
\dot x_2&=&-\delta x_2-k_2x_3x_2+s_2 \label{model2} \\
\dot x_3&=&-\gamma_0 x_3+u_2 \label{model3} \\
\dot x_4&=&g\dfrac{x_1}{h+x_1}x_4-rx_4-p_0x_4x_1-
k_1x_4 x_3+s_1u_1 \label{model4}
\end{eqnarray}  
where
\begin{tabbing}
\hskip 1cm \= \hskip 6cm \kill 
$x_1$ \> tumor cell population \\
$x_2$ \> circulating lymphocytes population\\
$x_3$ \> chemotherapy drug concentration \\
$x_4$ \> effector immune cell population \\
$u_1$ \> rate of introduction of immune cells \\
$u_2$ \> rate of introduction of chemotherapy 
\end{tabbing}
The description of the role of each groups of term is given in Table \ref{tab1} for the sake of clarity. 
\begin{table}[H]
\begin{center}
\begin{tabular}{lll} \toprule
    {Eq.} & Term & Description \\ \midrule
    (\ref{model1}) & $ax_1(1-bx_1)$ & {\footnotesize Logistic tumor growth}\\    
    (\ref{model1})  & $-c_1x_4x_1$ & {\footnotesize Death of tumor due to effector cells}\\
    (\ref{model1})  & $-k_3x_3x_1$ & {\footnotesize Death of tumor due to chemotherapy}\\
    (\ref{model2})  & $-\delta x_2$ & {\footnotesize Death of circulating lymphocytes}\\
(\ref{model2})   & $-k_2x_3x_2$ & {\footnotesize Death of lymphocytes due to chemo}\\
(\ref{model2})   & $s_2$ & {\footnotesize Constant source of lymphocytes}\\
(\ref{model3})   & $-\gamma_0 x_3$ & {\footnotesize Exponential decay of chemotherapy}\\    
(\ref{model4})   & $g\dfrac{x_1}{h+x_1}x_4$ & {\footnotesize Stimulation of tumor on effector cells}\\
(\ref{model4})   & $-rx_4$ & {\footnotesize Death of effector cells}\\
(\ref{model4})   & $-p_0x_4x_1$ & {\footnotesize Inactivation of effector cells by tumor}\\    
(\ref{model4})   & $-k_1x_4x_3$ & {\footnotesize Death of effector cells due to chemo}\\    
\bottomrule
\end{tabular}
\end{center} 
\ \\
\caption{Signification of the terms involved in the dynamic model (\ref{model1})-(\ref{model4}) [source \cite{Kassara2011135}]} \label{tab1} 
\end{table}

When referred to the nominal values of the parameters involved in the model (\ref{model1})-(\ref{model4}), the following values are used:
{\footnotesize
\begin{align*}
&a = 0.25, b = 1.02\times 10^{-14}, c_1 = 4.41 \times 10^{-10},g = 1.5\times 10^{-2}\\ 
&h = 20.2, k_2=k_3=0.6, k_1 = 0.8, p_0 = 2\times 10^{-11}, r = 0.04
\\&s_1 = 1.2\times 10^7, s_2 = 7.5\times 10^6, \delta = 1.2\times 10^{-2}, \gamma_0 = 0.9
\end{align*}}

Note that the dynamic model (\ref{model1})-(\ref{model4}) involves $14$ parameters. It is worth noting however that the order of magnitude of $x_1$ is about $10^9$ while $h\approx 20$. This means that $h$ acts only when the tumor is almost disappearing\footnote{On the other hand, without $h$, the effector immune population will keep growing while the tumor is absent which contradicts its very definition as a tumor stimulated immune population.}. That is the reason why in the remainder of this paper, this parameter is supposed to be known and hence it is not included in the set of uncertain parameters. As it is shown later, this enables a separable structure of the evolution map to be used [see (\ref{dynamicsPhiPsi})].

Gathering all the other parameters in a vector $p\in \mathbb{R}^{13}$ and using some discretization scheme with some sampling period $\tau$, it is possible to put the dynamics above in the following condensed form for the easiness of notation:
\begin{align}
x^+&=f(x,u,p)\quad (x,u,p)\in \mathbb{R}^{4}\times \mathbb{R}^{2}\times \mathbb{R}^{13} \label{dynamics}  \\
&=\Phi(x,u)\Psi(p)\label{dynamicsPhiPsi}
\end{align}
for straightforward definition of $\Phi$ and $\Psi$. 

The trajectory of the system starting at initial state $x$ under a control profile $\bm u$ and when the parameter value $p$ holds is denoted by $x^{\bm u}(k\vert x,p)$, namely:
\begin{align*}
&x^{\bm u}(k+1\vert x,p) = f\bigl(x^{\bm u}(k\vert x,p),u(k),p\bigr)\\
&x^{\bm u}(0\vert x,p)=x
\end{align*}

\subsection{The control objective and constraints}
The aim of the combined therapy is to reduce the size of the tumor population $x_1$ while keeping the level of lymphocyte cells $x_2$ above an upper bound $x_2^{min}$. This has to be done using quantized drug delivery:
\begin{equation}
u\in \mathbb U:=\{0,u_1^{max}\}\times \{0,u_2^{max}\}
\end{equation} 
so that an infinite sequence of control lying inside $\mathbb U$ is denoted hereafter by $\mathbb U^\infty$.

In order to address the control objective, the following stage cost function can be used that incorporates the constraint on $x_2$ as a soft constraint:
\begin{equation}
L(x,u):=x_1^2+\rho_c\max\{0,x_2^{min}-x_2\}+\rho_1u_1+\rho_2u_2
\end{equation}
so that {\bf  if the model's parameters were perfectly known},  an optimization problem can be defined for all given initial state $x$ as follows:
\begin{align}
&\mathcal P(x\vert p):\nonumber \\
&\quad \min_{\bm u\in \mathbb U^\infty}J(\bm u\vert x, p):=\sum_{k=0}^{\infty} \gamma^k L\Bigl(x^{\bm u}(k\vert x,p),u(k)\Bigr) \label{defdeJnom}
\end{align}
for some $\gamma\in (0,1]$. Now, since the parameters are supposed to be unknown, the stochastic control approach amounts at replacing the deterministic cost function (\ref{defdeJnom}) by a statistically defined one such as:
\begin{align}
J_\alpha(\bm u\vert x,\Pi)&:= \mu(J(\bm u\vert x,\cdot))+\alpha \sigma(J(\bm u\vert x,\cdot)) \label{defdealpha}\\
&=: \mathcal S_\alpha^{(J)}(\bm u,x) \label{defdecalS}
\end{align}
where $\mu(\cdot)$ and $\sigma(\cdot)$ stand respectively for the expectation and the variance of their argument when $p$ is sampled according to some supposedly known Probability Density Function (pdf) $\Pi$. Note that (\ref{defdecalS}) simply introduces a short notation of the r.h.s of (\ref{defdealpha}).

 This leads to the following stochastic optimization problem:
\begin{equation}
\mathcal P_\alpha(x\vert \Pi): \quad \min_{\bm u\in \mathbb U^\infty}J_\alpha(\bm u\vert x, \Pi) \label{defdeJstoch}
\end{equation}
Note that in many stochastic control-related works, attention is focused on the expectation of the cost function so that $\alpha=0$ is widely used in (\ref{defdealpha}). This is not totally appropriate in the case of cancer therapy. Indeed, expectation is a relevant indicator only when high number of realizations are expected to take place\footnote{such as saving energy in building management for instance despite of bad knowledge of exogenous parameters such as power demands and whether conditions.} in which only the global average is of interest. In the case of cancer therapy, a scenario is a patient being treated. This is why one should obviously be interested in the risk that {\bf  each patient afford} during the treatment. In such cases, it is precisely those {\em bad} scenarios with non negligible likeliness that really matter.  This has to do not only with the expectation but definitively with the associated variance as well. This explains the use of $\alpha>0$ in (\ref{defdeJstoch}). 

A part of the discussion and a large part of the results shown in this paper are dedicated to showing the relevance of introducing the variance related term the cancer therapy stochastic control problem's formulation on one hand, and on the convergence of the computation scheme in the presence of a non vanishing $\alpha$ in the stochastic cost formulation. 

In the next section, some recalls are proposed on the Stochastic Dynamic Programming equations that might be invoked to approximately solve the stochastic optimization problem (\ref{defdeJstoch}). 

\section{Recalls and preliminary results} \label{SecRecalls}
\subsection{Stochastic Dynamic Programming}
SDP attempts to solve the following functional equation in which the unknown function $V(\cdot)$ is the optimal solution of (\ref{defdeJstoch}). More precisely:
\begin{align}
V(x)&:= \min_{u\in \mathbb U} Q(x,u)\qquad \mbox{\rm  where} \label{defdeVdex} \\
Q(x,u)&:= L(x,u)+\gamma\min_{v\in \mathbb U}\Bigl[\mathcal S_\alpha^{(Q)}(x,u,v)\Bigr] \label{defdeQxu}\\
& \mbox{\rm where $x^+=f(x,u,p)$}
\end{align}
in which $\mathcal S^{(Q)}(x,u,v)$ is defined in a similar way as $\mathcal S^{(J)}(\bm u,x)$ [see (\ref{defdecalS})]:
\begin{equation}
\mathcal S^{(Q)}_\alpha(x,u,v) := \mu\bigl(Q(f(x,u,\cdot),v)\Bigr) +\alpha\sigma\bigl(Q(f(x,u,\cdot),v)\Bigr) \label{fedre}
\end{equation}
where here again, $\mu(\cdot)$ and $\sigma(\cdot)$ are respectively the expectation and the variance of their argument when $p$ (involved in the definition of $f(x,u,p)$) is sampled according to the pdf $\Pi$. 

The fact that $V$ satisfying (\ref{defdeVdex})-(\ref{defdeQxu}) is a solution to (\ref{defdeJstoch}) is a direct consequence of the Bellman principle and the fact that $L=\mu(L)+\sigma(L)=L+0$ since $L(x,u)$ is a deterministic $p$-unrelated term. 

From the above, it comes out that the truly unknown function to find is $Q(\cdot,\cdot)$ since the optimal value function $V(\cdot)$ as well as the optimal control are then recovered from the static optimization problem (\ref{defdeVdex}).

Now it is not hard to see from (\ref{defdeQxu}) that $Q$ is a solution of a fixed-point iteration:
\begin{equation}
Q^{(i+1)}=F\Bigl(Q^{(i)}\Bigr) \label{fixedpoint}
\end{equation}
where the operator $F$ (commonly called the Bellman operator) is defined by:
\begin{equation}
F(Q)(x,u):= L(x,u)+\gamma\min_{v\in \mathbb U}\Bigl[\mathcal S^{(Q)}(f(x,u,\cdot),v)\Bigr] \label{FPvraiment}
\end{equation}

In the next section, some convergence results regarding the above fixed-point iterations are derived in a general conceptual setting before a specific implementation is proposed in Section \ref{SecPM} for the cancer combined therapy. 
\subsection{Fixed-Point iteration convergence analysis}
In order to derive sufficient conditions for the convergence of the fixed-point iteration (\ref{fixedpoint}), some definitions and assumptions are needed. They are introduced hereafter. 

First of all, the following continuity assumption of the dynamics is used:

\begin{hyp}[Continuous dynamics]\label{hyp1}
The map $f$ describing the discrete-time state evolution (\ref{dynamics}) is continuous in its arguments. 
\end{hyp}

Regarding the uncertain vector $p$, the following assumption is needed
\begin{hyp}[Finitely supported uncertainties] \label{hyp2}
there exists a compact set $\mathbb P$ to which belong all possible realizations of the uncertain vector $p$.
\end{hyp}

This seems a quite reasonable assumption when physically meaningful model's parameters are invoked. Still, this is rigorously not the case for Gaussian distributions for instance for which any arbitrarily high values are rigorously admissible. 
\begin{definition}[Bounded excursion]\label{defbound}
A map $G$ defined on $\mathbb X\times \mathbb U$ is said to have $B$-{\em bounded excursions} if the following inequality holds:
\begin{align}
\vert G(f(x,u,p),v)-\mu(G(f(x,u,\cdot),v))\vert \le B
\end{align}
for all $(x,u,v,p)\in \mathbb X\times \mathbb U^2\times \mathbb P$. 
\end{definition}
This definition simply states that the excursion of the realizations $G(f(x,u,p),v)$ from the mean value $\mu\bigl(G(f(x,u,\cdot),v)\bigr)$ is bounded. It goes without saying that, under Assumption \ref{hyp1}, such a bound exists as soon as the map $G$ is continuous. since all the involved bounding sets are compact. 

\begin{lem}[Fixed-Point convergence] \label{lemma1}
Under Assumptions \ref{hyp1} and \ref{hyp2}, if the fixed point map $F$ invoked in (\ref{fixedpoint}) produces only {\bf continuous} maps $Q^{(i)}$ with $B$-bounded excursions for some $B>0$, then the fixed point iteration converges on $\mathbb Z:= \mathbb X\times \mathbb U$ provided that the penalty $\alpha$ on the variance term in (\ref{fedre}) satisfies:
\begin{equation}
\alpha < \dfrac{\gamma-1}{2\gamma B} \label{condonalpha}
\end{equation}
and for sufficiently small sampling period $\tau$ that is used to derive the discrete-time dynamics (\ref{dynamics}).
\end{lem}
{\sc Proof}. See Appendix \ref{appendLemma1}.

From the above lemma, one can clearly derive the standard convergence result that is known to hold when no penalty on the variance is used (see for instance \cite{Jaakoola1994,Kirkby2017,Bertsekas2017} and the references therein):
\begin{cor}[Convergence for pure expectation]\label{corpure}
Under the assumptions of Lemma \ref{lemma1}, if the cost is defined exclusively on the expectation (no penalty on the variance in (\ref{fedre})), then the fixed point iteration is convergent on $\mathbb Z=\mathbb X\times \mathbb U$. 
\end{cor}
It is worth underlying that the convergence results cited above depend on the assumption according to which the successive iterates $Q^{(i)}$ produced by the Bellman map $F$ admit a $B$-bounded excursions with common upper bound $B$. The way this condition is enforced is discussed in the next section. 

\section{The proposed computation framework} \label{SecPM}

\subsection{Parametrization of $Q(x,u)$} \label{recallML}
In the previous section, the SDP equation and the related fixed-point iteration have been defined in a conceptual framework. In this framework, the unknown function $Q(x,u)$ one is looking for has no specific finite-dimensional representation that would be compatible with concrete computation schemes. In this section, this parameterization aspect is addressed. 

For the sake of simplicity, the notation $z=(x,u)$ and $\mathbb Z:=\mathbb X\times \mathbb U$ is used so that $Q(z)$, $f(z,p)$ stands for $Q(x,u)$ and $f(x,u,p)$ respectively.  The finite dimensional parameterization of $Q$ is obtained through the following steps:
\begin{itemize}
\item First of all, a fixed grid of points $\mathcal Z:=\{z^{(i)}\}_{i=1}^{n_g}$ is defined on the compact subset $\mathbb Z$ of interest. \\
\item Denote by $\bm q:=\{q_i\}_{i=1}^{n_g}$ the vector of values of the function $Q$ at the grid points, namely $q_i=Q(z^{(i)})$.\\
\item Choose some regression model (Polynomial, Gaussian, Support Vector Machine  Regressor (SVMR), etc.) and denoted it by $\hat Q$. As far as this work is concerned, a SVMR with gaussian basis is used with the default parameters used in the scikit-learn Machine Learning (ML) library \cite{scikit-learn}. \\ 
\item This choice enables to associate to each vector of values $\bm q\in \mathbb{R}^{n_g}$ a function $\hat Q(\cdot\vert \bm q)$ defined on $\mathbb Z$ which is precisely the one identified (or {\em learned} to use a ML vocabulary) using the learning data $(\mathcal Z,\bm q)$. 
\end{itemize}
Based on the above items, it is now possible to state the following definition:
\begin{definition}[Admissible regression model]\label{adregmod}
A regression model is said to be admissible if and only if for all set $\mathcal Z$ of finite grid points in $\mathbb Z$ and any compact set of function values $\mathbb Q\subset \mathbb{R}^{n_g}$, there exists $B>0$ such that all identified maps $\hat Q(\cdot\vert \bm q)$ with $\bm q\in \mathbb Q$ shows $B$-bounded excursions in the sense of Definition \ref{defbound}.
\end{definition}
\subsection{Main convergence result} \label{subsecmain}
Using the above preliminary results and definitions, the following main result regarding the convergence of the proposed Fixed-Point iteration scheme can be stated as follows:
\begin{prop}[Main result] \label{propmain}
Under Assumptions \ref{hyp1} and \ref{hyp2}, if an admissible regression model(in the sense of Definition \ref{adregmod}) is used  over some grid points $\mathcal Z:=\{z^{(i)}\}_{i=1}^{n_g}$ with the initialization $q_i=L(z^{(i)})$ to solve the stochastic optimal feedback of the combined therapy of cancer as stated in Section \ref{SecPs}, then the resulting fixed-point iteration (\ref{FPvraiment}) with $\gamma\in (0,1)$ is convergent provided that the variance related penalty $\alpha$ and the sampling period $\tau$ are taken sufficiently small.
\end{prop}
{\sc  Proof}. See appendix \ref{appmainresult}. 

The only remaining task is to find a regression model that is admissible in the sense of Definition \ref{adregmod}. Fortunately, there are many options. The one that is used in the remainder of this paper is the Support Vector Machine regressor (SVMR) which shows the following structure:
\begin{equation}
\hat Q(z\vert\bm q):=\beta_0(\bm q)+\sum_{i=1}^{n_g}\beta_i(\bm q) G(z^{(i)},z)
\end{equation}
where the coefficients $\beta_i(\bm q)$ admits uniform bound if $\bm q$ belongs to some compact set $\mathbb Q$. 
\begin{rem}
It is worth mentioning at this stage that the convergence of the fixed point-iteration does not tell us much on the distance of the resulting solution to the true optimal solution. This heavily depends on the density of the grid $\mathcal Z$ on one hand and on the quality of the approximation of the statistical quantities involved in the very definition of the cost function on the other hand. These two aspects remain obviously at the origin of the {\em curse of dimensionality} that restricts the application of the underlying SDP approach to system of moderate sizes.
\end{rem}
\subsection{Computation of the expectation and variance} \label{statissec}
The fixed-point iteration (\ref{FPvraiment}) requires the computation of the map $\mathcal S^{(Q)}(f(x,u,\cdot),v)$ which involves the computation of the expectations and variances of quantities over the probability space of parameter values. In this section, the way this is done concretely is explained. 

Recall that the r.h.s of the dynamics in (\ref{dynamicsPhiPsi}) can be decomposed into two multiplicative terms $x^+=\Phi(x,u)\Psi(p)$ where only the second term is parameter-dependent. Consequently, the computation of the expectation of $x^+$ needs the statistics of $\Psi(p)\in \mathbb{R}^{14}$ to be approximated. This is done by identifying $n_{cl}$ clusters based on a set $\mathcal S_\psi$ of $n_s$ samples of the vector $\Psi(\cdot)$ that are drawn using the pdf $\Pi$. The centers of the computed clusters, denoted by $\Psi^{(j)}$, $j=1,\dots,n_{cl}$ can be used as representatives of their clusters. Moreover, the ratio $\pi_j$ between the sizes of these clusters and the total population $n_s$ can be used as a measure of their probabilities, namely
\begin{equation}
\pi_j:=\dfrac{1}{n_s}\mbox{\rm card}\Bigl\{\psi\in \mathcal S_\psi\quad\vert\quad  \psi=\Psi^{(j)}\Bigr\}
\end{equation}

Based on the above definitions, the following approximation are used in the approximation of terms in (\ref{fedre}):
\begin{align}
&\hat\mu\bigl(x,u,v\vert \bm q\bigr):= \sum_{j=1}^{n_{cl}}\pi_j\hat{Q}\Bigl(\Phi(x,u)\Psi^{(j)},v \vert \bm q\Bigr) \label{computemu}\\
&\hat\sigma\bigl(x,u,v,\vert \bm q\bigr):= \nonumber \\ &\sum_{j=1}^{n_{cl}}\pi_j\Biggl[\hat Q\Bigl(\Phi(x,u)\Psi^{(j)},v\vert \bm q\Bigr)-\hat\mu\bigl(x,u,v\vert \bm q\bigr)\Biggr]^2 \label{computemu}
\end{align}
Using these expression, the weighted sum (\ref{fedre}) of the above term can be used to approximated by:
\begin{equation}
\hat{\mathcal S}_{\alpha}^{(Q)}(x,u,v\vert \bm q):= \hat\mu\bigl(x,u,v\vert \bm q\bigr)+\alpha \hat\sigma\bigl(x,u,v\vert \bm q\bigr)
\end{equation} 
The updated values of $q_i$ at the underlying grid points $z^{(i)}$ can therefore be obtained through:
\begin{equation}
q^+_{i}:=L(z^{(i)})+\gamma\min_{v\in \mathbb U}\Bigl[\hat{\mathcal S}_\alpha^{(Q)}(f(z^{(i)}),v\vert \bm q)\Bigr] 
\end{equation}
for each element $z^{(i)}$ of the underlying grid $\mathcal Z$. Note that the optimization problem in $v$ can be solved by simple enumeration over the four elements of the set $\mathbb U$.

This implements a fixed-point iteration on the vector of values $\bm q$ than can be shortly written as follows:
\begin{equation}
\bm q^+=F(\bm q) \label{Fdeq}
\end{equation}
where the notation $F$ used in (\ref{fixedpoint}) is overloaded using the finite-dimensional parametrization $\bm q$ of $Q$. This is the fixed-point iteration that is proved to be convergent under the assumptions of Proposition \ref{propmain}. 

In Section \ref{SecRes}, the instantiation of the framework on the combined therapy of cancer is detailed and the corresponding results are commented. 

\section{Numerical investigation on the combined therapy of cancer} \label{SecRes}
\subsection{Parameters used in the numerical investigation} \label{numparam}
As far as stochastic controller design is concerned (nominal controller is also investigated), the above framework has been investigated using the following parameters. A number $n_s=10^4$ samples have been drawn to perform the clustering of set of values of $\Psi$ (see Section \ref{statissec}). The number of identified clusters was taken equal to $n_{cl}=20$. 

Regarding the pdf $\Pi$, the following definition has been used $p_i=(1+\nu)p_i^{nom}$ where $\nu$ is a normal distribution with $0$ mean and variance $\sigma=0.4$. The grid $\mathcal Z$ has been taken as the cartesian product of a uniform grid on $x$ (having $7$ uniformly distributed values over the normalized interval $[0,1]$ per component) and the four values control set $\mathbb U$. This induces a set $\mathcal Z$ composed of $7^4\times 4=9604$ elements. 

Clustering has been performed using the scikit-learn library's {\sc KNN} (k-nearest neighbors) utility. The scikit learn library's SVM regression model has been used with radial basis function's kernel and the default values of the remaining parameter. The sampling period $\tau=0.25$ Day is used and the lower bound $x_2^{min}=0.05$ is used for the normalized\footnote{The state components have been normalized using respectively the reference values $(10^9,10^9,1,10^9)$.} lymphocyte population size. 

Three different controllers are compared. Namely:
\begin{enumerate}
\item The nominal controller for which $\Pi$ is a dirac distribution centered at the nominal vector of parameters. The is referred to as Controller \#1.\\
\item The expectation-based controller for which the distribution of the parameter is the one described above \ref{numparam} where no penalty on the variance term ($\alpha=0$). The is referred to as Controller \#2.\\
\item The mixed expectation/variance-based controller for which the penalty $\alpha=0.1$ is used on the variance-related term. The is referred to as Controller \#3.
\end{enumerate}
\subsection{Results}
Despite the fact that the convergence of the fixed-point iteration has been proved only for $\gamma\in (0,1)$ and for sufficiently small $\alpha$ and $\tau$, the value $\gamma=1$ has been used to show that convergence does occur for this ideal value (no discount on the cost function). On the other hand $\alpha=0.1$ for Controller \#3. 

\begin{figure}[H]
\begin{center}
\includegraphics[width=0.5\textwidth]{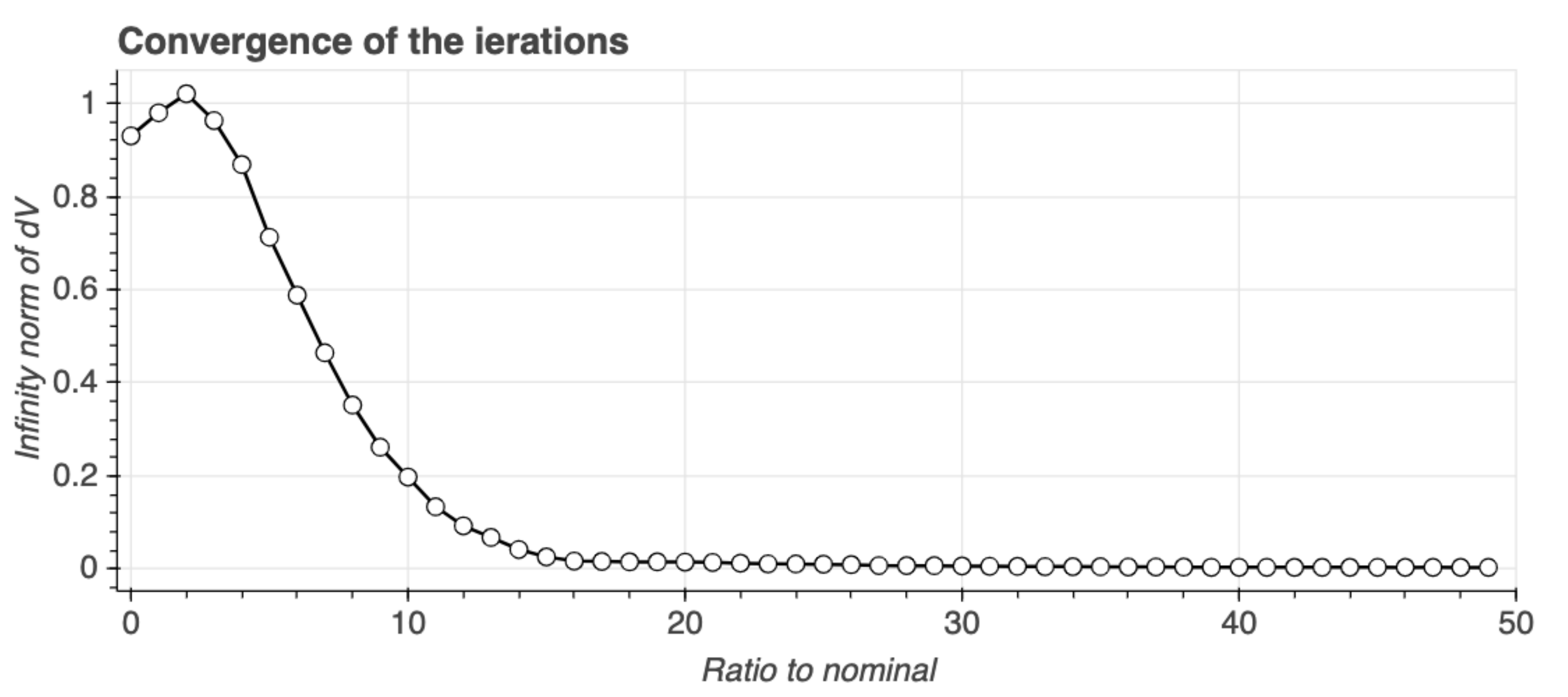}
\end{center}
\caption{Typical convergence curve for fixed-point iterations: Evolution of the difference $\|\bm q^{(i+1)}-\bm q^{(i)}\|_\infty$ between two successive iterations. $\gamma=1$, $\alpha=0.1$.} \label{figConvergenceFP}
\end{figure}

\begin{figure}[H]
\begin{center}
\includegraphics[width=0.5\textwidth]{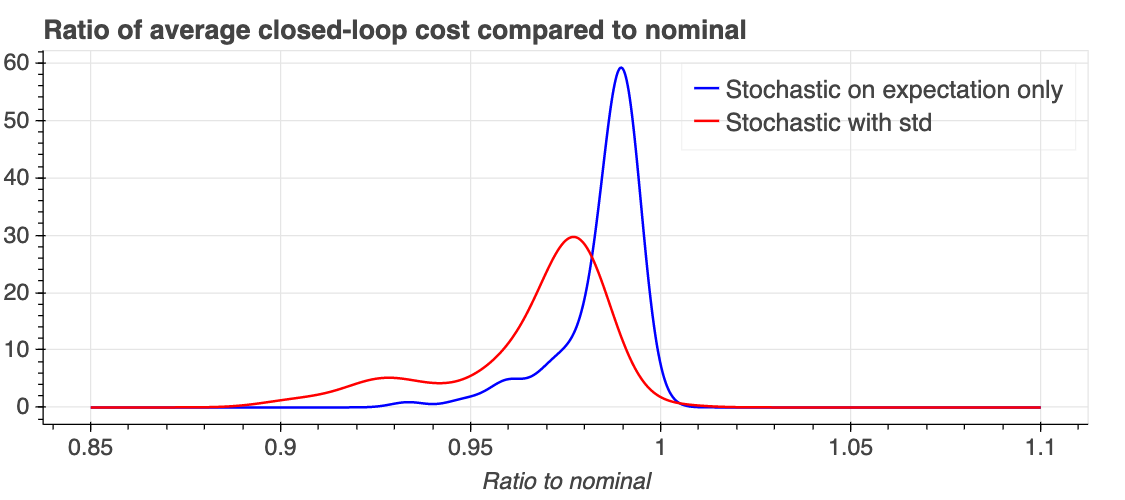}
\end{center}
\caption{Comparison between the closed-loop performance ratios relatively to the nominal controller for the two stochastic formulations for the 20000 randomly sampled initial states and parameter vectors.} \label{figCLCostComparison}
\end{figure}

\begin{figure}[H]
\begin{center}
\includegraphics[width=0.5\textwidth]{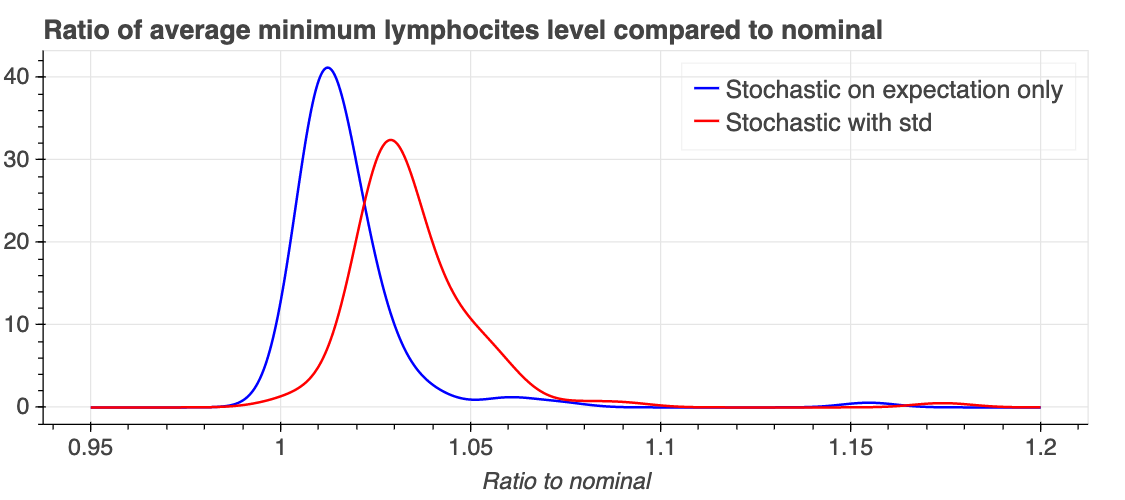}
\end{center}
\caption{Comparison between the  average lymphocytes population level ratios relatively to the nominal controller for the two stochastic formulations for the 20000 randomly sampled initial states and parameter vectors.} \label{figConstraintsComparison}
\end{figure}

First of all, Figure \ref{figConvergenceFP} shows a typical convergence curve for the fixed-point iteration. The difference between two successive solutions during the iterations, namely $\|\bm q^{(i+1)}-\bm q^{(i)}\|_\infty$
 is shown. Note that because of the use of $\gamma=1$, no strict contraction is obtained but eventually, contraction is settled and the solution is obtained. 
Figures  \ref{figCLCostComparison} and \ref{figConstraintsComparison} show the benefit from using the stochastic formulation compared to the nominal one. In order to show this, $N_{sam}:=20000$ different simulations are constructed by drawing randomly samples of pairs of initial states and parameter vectors and the resulting closed-loop cost function is simulated before statistics are drawn. More precisely, $100$ initial states are sampled and for each of these initial state, $200$ random parameter vectors samples are drawn. Namely, for each given pair $(x^{(i)},p^{(i)})$, $i=1,\dots,N_{sam}$
\begin{equation}
J_{cl}(x^{(i)},p^{(i)}):=\sum_{k=0}^{N_{sim}} L(x(k),u(k)) \label{clperf}
\end{equation}
where $N_{sim}=50$ is the simulation horizon length (12.5 Days) while $x(k)$ and $u(k)$ stand for the closed-loop trajectories at instant $k$ starting at initial instant $k=0$ at $x^{(i)}$ and when the parameter vector $p^{(i)}$ is used. Then different statistical comparisons are done when the above procedure is executed using the three controllers mentioned above, namely, the nominal, the stochastic controller based on the expectation term only ($\alpha=0$) and finally the stochastic controller using a penalty $\alpha=0.1$ on the variance. 

More precisely, Figure \ref{figCLCostComparison} shows the histogram of the ratios between the closed-loop performance as defined in (\ref{clperf}) when the stochastic controllers are used compared to the nominal. One can clearly observe two features:
\begin{enumerate}
\item Both stochastic controllers lead to statistically better results than the nominal. \\ 
\item The stochastic Controller \#3 that incorporates penalty on the variance shows better results than the one using only the expectation-related 
penalty term. \\
\item It seems clear that even when focusing on the mean value of the cost function, the controller \# 3 shows a better average than the controller \# 2. This might seem contradictory with the fact that the latter is supposed to minimize this expectation. It can be conjectured that this comes from the fact that the expectation cannot be sufficiently well represented by the average over the $20$ clusters (see Section \ref{statissec}) that are used to induce tractable computation of the expectation and the variance and considering the penalty on the variance correct this bias in a favorable way. 
\end{enumerate}
Figure \ref{figConstraintsComparison} shows the comparison between the average levels of the lymphocytes population cells sizes (compared to the nominal) when the stochastic controllers are used. Here again, it comes out that the performances of the controller \# 3 are slightly bette than that of the second expectation-only-based controller. 
\begin{figure*}[t]
\begin{center}
\includegraphics[width=0.85\textwidth]{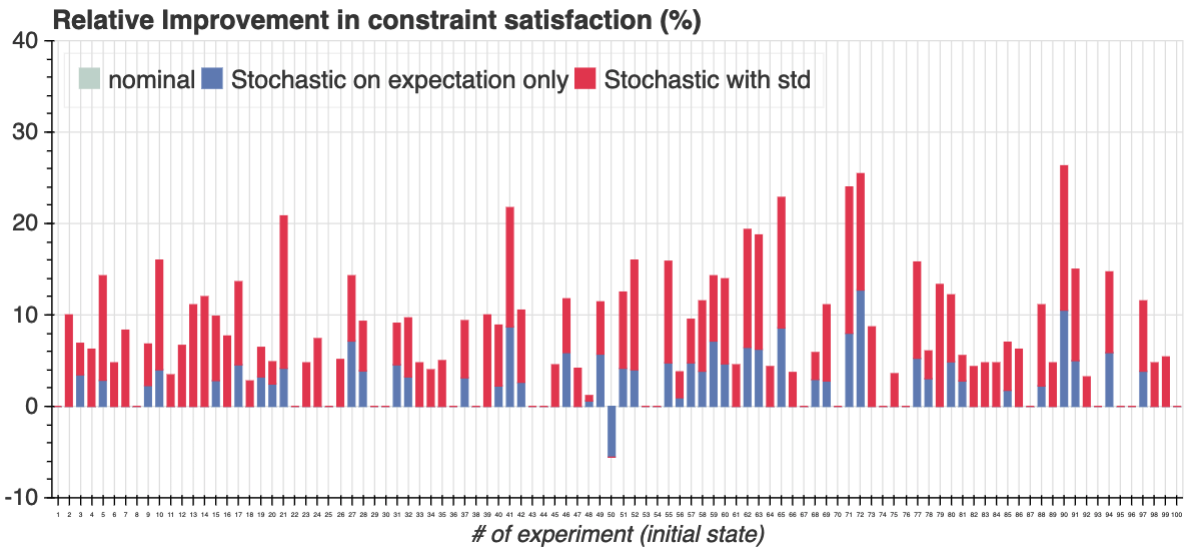}
\end{center}
\caption{Safety constraints satisfaction improvement relatively to the nominal controller for the two stochastic formulations. Abscissa represents the initial state number (100 are randomly selected) and the percentage refers to the $200$ parameter vectors that are simulated for the same initial state. } \label{figBarComparison}
\end{figure*}

\noindent Another way to look at the constraint satisfaction is to count the number of scenarios where the health constraint $x_2\ge x_2^{min}$ is satisfied. Figure \ref{figBarComparison} shows the percentage of improvement in the constraints satisfaction compared to the nominal when the stochastic controllers are used. 
\section{Conclusion and future works} \label{SecConc}
In this paper a tractable stochastic control design for the combined therapy of cancer is proposed. The method is based on an approximate solution of the SDP equations through a value function fixed-point iteration. The convergence of the latter in the presence of variance related penalty in the cost function is proved under mild technical assumption. The result suggests that the inclusion of a variance-related term in the cost function might give better result by partially compensating for the errors induced by the approximation of the statistical quantities through computations involving a reduced number of clusters. Undergoing work involves the use of GPU-based computation framework to handle a higher order of magnitude of the samples size in order to examine how these qualitative results scale with the dimension of the populations over which they are evaluated. 

\bibliography{biblio_cancer_scenarios}
\bibliographystyle{plain}

\appendix
\section{Appendix}
\subsection{Proof of Lemma \ref{lemma1}} \label{appendLemma1}
{\sc  Proof}. 
We shall prove that under the assumptions of the Lemma, the following inequality holds on the fixed-point map $F$
\begin{equation}
\|F(Q_1)-F(Q_2)\|_\infty \le \gamma^{'}\|Q_1-Q_2\|_\infty
\end{equation}
where 
\begin{equation}
\|Q\|_\infty = \sup_{(x,u)\in \mathbb \mathbb X\times \mathbb U} \vert Q(x,u)\vert 
\end{equation}
for some $\gamma^{'}\in [0,1)$ which would obviously be sufficient to prove the result. To do so, let us begin by deriving upper bounds on the norm of $\vert F(Q_1)(x,u)-F(Q_2)(x,u)\vert $. Note that one has by definition of $F$:
\begin{equation}
[F(Q_1)-F(Q_2)](x,u):=\gamma  \Bigl[G_\mu(x,u)+\alpha G_\sigma(x,u)\Bigr] \label{FQ1mFQ2}
\end{equation}
where $G_\mu$ comes from the mean-related term in (\ref{fedre}):
\begin{align}
&G_\mu(x,u):=\label{Gmu} \\
&\mu\Bigl[Q_1(f(x,u,\cdot),v_1^\dag)\Bigr]\nonumber 
-\mu\Bigl[Q_2(f(x,u,\cdot),v_2^\dag)\Bigr]\nonumber 
\end{align}
where $v_1^\dag$ and $v_2^\dag$ are the solution of the corresponding optimisation problems. Similarily:
$G_\sigma$ is induced by the variance-related term in (\ref{fedre}), namely:
\begin{align}
&G_\sigma(x,u):=\label{Gsigma} \\
&\mu\Biggl[\Bigl(Q_1(f(x,u,\cdot),v_1^\dag)-\mu\bigl[Q_1(f(x,u,\cdot),v_1^\dag)\bigr]\Bigr)^2\Biggr]- \nonumber \\
&\mu\Biggl[\Bigl(Q_2(f(x,u,\cdot),v_2^\dag)-\mu\bigl[Q_2(f(x,u,\cdot),v_2^\dag)\bigr]\Bigr)^2\Biggr]\nonumber
\end{align}
Let us consider successively upper bounds on $\vert G_\mu\vert $ and $\vert G_\sigma\vert$.

\underline{Bounding $\vert G_\mu(x,u)\vert$}:

By using the pdf $\Pi$, it is possible to write:
\begin{align*}
&\vert G_\mu(x,u)\vert \le  \\
&\int \Pi(p)\left\vert Q_1(f(x,u,p),v_1^\dag)-Q_2(f(x,u,p),v_2^\dag)\right\vert dp
\end{align*}
Now since by definition:
\begin{equation*}
(f(x,u,p),v^\dag_i)\in (1+O(\tau))\mathbb X\times \mathbb U
\end{equation*}
it comes from the previous equation that:
\begin{align}
\vert G_\mu(x,u)\vert\le& (1+O(\tau))\int \Pi(p)\|Q_1-Q_2\|_\infty dp\\ 
=&(1+O(\tau))\|Q_1-Q_2\|_\infty \label{enfinGmu}
\end{align}
\underline{Bounding $\vert G_\sigma(x,u)\vert$}\\ \ \\ 
one can write after straightforward manipulations (mainly using the identity $a^2-b^2=(a-b)(a+b)$):
\begin{align*}
&G_\sigma(x,u) = \\
& \int \Pi(p)\Bigl[Q_1(f(x,u,p),v_1^\dag)-Q_2(f(x,u,p),v_2^\dag)- \\
& \mu\bigl(Q_1(f(x,u,\cdot),v_1^\dag)-Q_2(f(x,u,\cdot),v_2^\dag)\bigr)\Bigr]\times \\ 
&\Bigl[Q_1(f(x,u,p),v_1^\dag)+Q_2(f(x,u,p),v_2^\dag)- \\
& \mu\bigl(Q_1(f(x,u,\cdot),v_1^\dag)+Q_2(f(x,u,\cdot),v_2^\dag)\bigr)\Bigr]dp
\end{align*}
Therefore, 
\begin{align*}
&\vert G_\sigma(x,u)\vert \le \\
& \int \Pi(p)\Bigl[Q_1(f(x,u,p),v_1^\dag)-Q_2(f(x,u,p),v_2^\dag)\Bigr]\times \\
&\vert\Bigl[Q_1(f(x,u,p),v_1^\dag)+Q_2(f(x,u,p),v_2^\dag)- \\
& \mu\bigl(Q_1(f(x,u,\cdot),v_1^\dag)+Q_2(f(x,u,\cdot),v_2^\dag)\bigr)\Bigr]\vert dp \\
&+ \\
& \vert \mu\bigl(Q_1(f(x,u,\cdot),v_1^\dag)-Q_2(f(x,u,\cdot),v_2^\dag)\bigr)\vert \times \\
& \vert \int \Pi(p) \Bigl[Q_1(f(x,u,p),v_1^\dag)+Q_2(f(x,u,p),v_2^\dag)- \\
& \mu\bigl(Q_1(f(x,u,\cdot),v_1^\dag)+Q_2(f(x,u,\cdot),v_2^\dag)\bigr)\Bigr]dp\vert
\end{align*}
Note that the second term in the above addition vanishes by definition of the mean of $Q_1+Q_2$. Regarding the first term, it can be bounded using the fact that $Q_1$ and $Q_2$ are both of finite excursion on $\mathbb X\times \mathbb U$. Indeed\begin{align*}
&\vert\Bigl[Q_1(f(x,u,p),v_1^\dag)+Q_2(f(x,u,p),v_2^\dag)- \\
& \mu\bigl(Q_1(f(x,u,\cdot),v_1^\dag)+Q_2(f(x,u,\cdot),v_2^\dag)\bigr)\Bigr]\vert \le 2B
\end{align*}
Using this in the first term of the last inequality bounding $\vert G_\sigma(x,u)\vert$ leads to
\begin{equation}
\vert G_\sigma(x,u)\vert \le 2B(1+O(\tau))\|Q_1-Q_2\|_\infty \label{enfinGsigma}
\end{equation}
Using (\ref{enfinGmu}) and (\ref{enfinGsigma}) in (\ref{FQ1mFQ2}) enables to induce that
\begin{equation}
\|F(Q_1)-F(Q_2)\|_\infty\le \gamma(1+2\alpha B+O(\tau))\|Q_1-Q_2\|_\infty
\end{equation}
which obviously means that for all $\alpha$ satisfying the condition (\ref{condonalpha}), the fixed point is contractive and hence convergent. $\hfill \Box$
\subsection{Proof of Proposition \ref{propmain}} \label{appmainresult}
All we have to prove is that when initializing the $q_i$ as stated in the proposition, all the following values of $\bm q$ lie in some compact set $\mathbb Q$. Once this is shown the results fallows directly from Definition \ref{adregmod} which guarantees that some unique $B$ exists which can be used in Lemma \ref{lemma1} to conclude. \\ \ \\ 
To prove the existence of the compact set $\mathbb Q$, the boundedness of the state evolution of the combined therapy of cancer is invoked, that is to say, under any bounded sequence $\bm u\in \mathbb U^{\infty}$, the state vector belongs at any instant in the future to some bounded set $\bar{\mathbb X}$ (this is a known property of the model (\ref{model1})-(\ref{model4}) that can be easily shown by proving that for each component of the state, the r.h.s becomes negative when the component is sufficiently high). 

Having this property at hand and given the initialization $q_i=L(z^{(i)}$, it comes from the definition of the fixed-point iteration that the values $q_i$ at any iteration satisfy the inequality:
\begin{equation}
q_i\le \dfrac{\gamma}{1-\gamma}\left[\max_{z\in \bar{\mathbb X}\times \mathbb U}L(z)\right]
\end{equation}
which obviously means that the fixed point iteration produces vectors that always remains in some compact set $\mathbb Q$ which ends the proof. $\hfill \Box$

\end{document}